\documentclass[reprint, superscriptaddress, secnumarabic, amssymb, nobibnotes, aps, prl]{revtex4-1}

\usepackage{graphicx}
\usepackage{epstopdf}
\usepackage[T1]{fontenc}
\usepackage[latin9]{inputenc}
\usepackage{amsbsy}
\usepackage{gensymb}
\begin{document}

\title{\textrm{Superconducting properties of the noncentrosymmetric superconductor LaPtGe}}
\author{Sajilesh.~K.~P.}
\affiliation{Department of Physics, Indian Institute of Science Education and Research Bhopal, Bhopal, 462066, India}
\author{D.~Singh}
\affiliation{Department of Physics, Indian Institute of Science Education and Research Bhopal, Bhopal, 462066, India}
\author{P.~K.~Biswas}
\affiliation{ISIS Facility, STFC Rutherford Appleton Laboratory, Harwell Science and Innovation Campus, Oxfordshire, OX11 0QX, UK}
\author{A.~D.~Hillier}
\affiliation{ISIS Facility, STFC Rutherford Appleton Laboratory, Harwell Science and Innovation Campus, Oxfordshire, OX11 0QX, UK}
\author{R.~P.~Singh}
\email[]{rpsingh@iiserb.ac.in}
\affiliation{Department of Physics, Indian Institute of Science Education and Research Bhopal, Bhopal, 462066, India}

\date{\today}
\begin{abstract}
\begin{flushleft}
\end{flushleft}
We report superconducting properties of noncentrosymmetric superconductor LaPtGe which crystallizes in noncentrosymmetric  $ \alpha $-ThSi$ _{2} $ structure. The magnetization, resistivity and specific heat measurements confirms that LaPtGe is a type-II bulk superconductor with a transition temperature T$ _{c} $ = 3.05 $ \pm $ 0.05 K. Muon-spin relaxation/rotation measurements confirms that time reversal preserved in the superconducting ground state.
\end{abstract}
\maketitle
\section{Introduction}

In the recent years, several new superconducting materials have been discovered. Many of these new superconductors are described as "unconventional", as their superconducting properties deviate from traditional BCS theory \cite{BCS}. Noncentrosymmetric superconductors (NCS) have emerged as an exciting class of new unconventional superconductors. The lack of inversion symmetry can make the pairing scenario different, which introduces the possibility of a vast array of exotic physics \cite{MAC,TP1,TP2,TP3}. A non-trivial antisymmetric spin orbit coupling (ASOC) arise due to an asymmetric electric field gradient lifts the original conduction electron spin degeneracy at the Fermi level, splitting it into two sublevels \cite{EBA}. This leads to an admix superconducting ground state which shows many exotic properties, which have not been observed in conventional superconductors \cite{rashba,sky,kv,ia,pa,fujimoto1,fujimoto2,fujimoto3,mdf}. The admixed pairing state can be manipulated by tuning the ASOC which directly controls the mixing ratio of singlet/triplet pairing channel \cite{LPB2}. In addition, the importance of electronic correlations cannot be neglected, which often facilitates the interaction between different pairing channels. Strongly correlated superconductors without inversion symmetry include CePt$ _{3} $Si \cite{Bauer2004}, Re$ _{6} $(Zr,Hf,Ti) \cite{rz3,rz,rhf,rh,rt,rf} and UIr \cite{UIr}, while LaNiC$ _{2}$ \cite{lnc2}, Li$ _{2} $M$ _{3} $B (M=Pd, Pt) \cite{LPB1,LPB2} and La$ _{7} $Ir$ _{3} $ \cite{LI} are  weakly correlated. Weakly correlated materials are of great importance since the effects of broken inversion symmetry and asymmetric spin-orbit coupling interactions can be more explicitly separated and understood.

At present, the major issue in this area is how the antisymmetric spin-orbit coupling influences the parity mixing in these materials and the presence/absence of time-reversal symmetry breaking (TRSB). Till now, only a small number of superconductors have been discovered that exhibit TRSB \cite{rz3,lnc2,LI,SRO,UB,UP} which makes it difficult to determine the roles of asymmetric spin-orbit coupling and electron correlations in non-centrosymmetric superconductors. Therefore, it is clearly crucial to search for new superconductors whose crystal structure lack inversion symmetry.

In this paper, we report a comprehensive study of the superconducting properties of the noncentrosymmetric ternary equiatomic compound LaPtGe, which is a ternary variant of $\alpha$-ThSi$ _{2} $ structure where a three-dimensional network of three connected metalloid atoms is found with tetragonal symmetry \cite{J}. Theoretical calculations on the ternary variant of the $\alpha$-ThSi$ _{2} $ compounds show strong spin-orbital coupling strength \cite{SOC}. F. Kneidinger et al. calculation on the parent ternary compound LaPtSi revealed strong ASOC \cite{LPS}. It is interesting to look for more compounds in the same family to find the effect of ASOC on the superconducting ground state, in particular LaPtGe in the present case. Here we have used resistivity, magnetization and heat capacity along with muon-spin spectroscopy to probe the gap symmetry and nature of the superconducting ground state.
\ 
\section{Experimental Details}
Polycrystalline LaPtGe samples were prepared using standard arc melting technique. High purity La (99.99\% ), Pt (99.99\%) and Ge (99.99\%) were taken in a stoichiometric ratio and melted in a water-cooled copper hearth under high purity Argon gas. The resulting ingot formed with the negligible mass loss was flipped and remelted several times to improve the homogeneity. Phase purity and crystal structure of the sample was confirmed by room temperature x-ray diffraction measurement using a PANalytical diffractometer equipped with CuK$\alpha$ radiation($\lambda$ = 1.5406 \AA). Magnetization measurements were done using a superconducting quantum interference device (MPMS 3, Quantum Design). The electrical resistivity and heat capacity measurements of the sample were performed using a Physical Property Measurement System (PPMS, Quantum Design). The $ \mu $SR experiments were carried out using a 100 \% spin-polarized pulse muon beam at ISIS facility, Rutherford Appleton Laboratory, Didcot, United Kingdom.
\begin{figure} 
\includegraphics[width=1.0\columnwidth, origin=b]{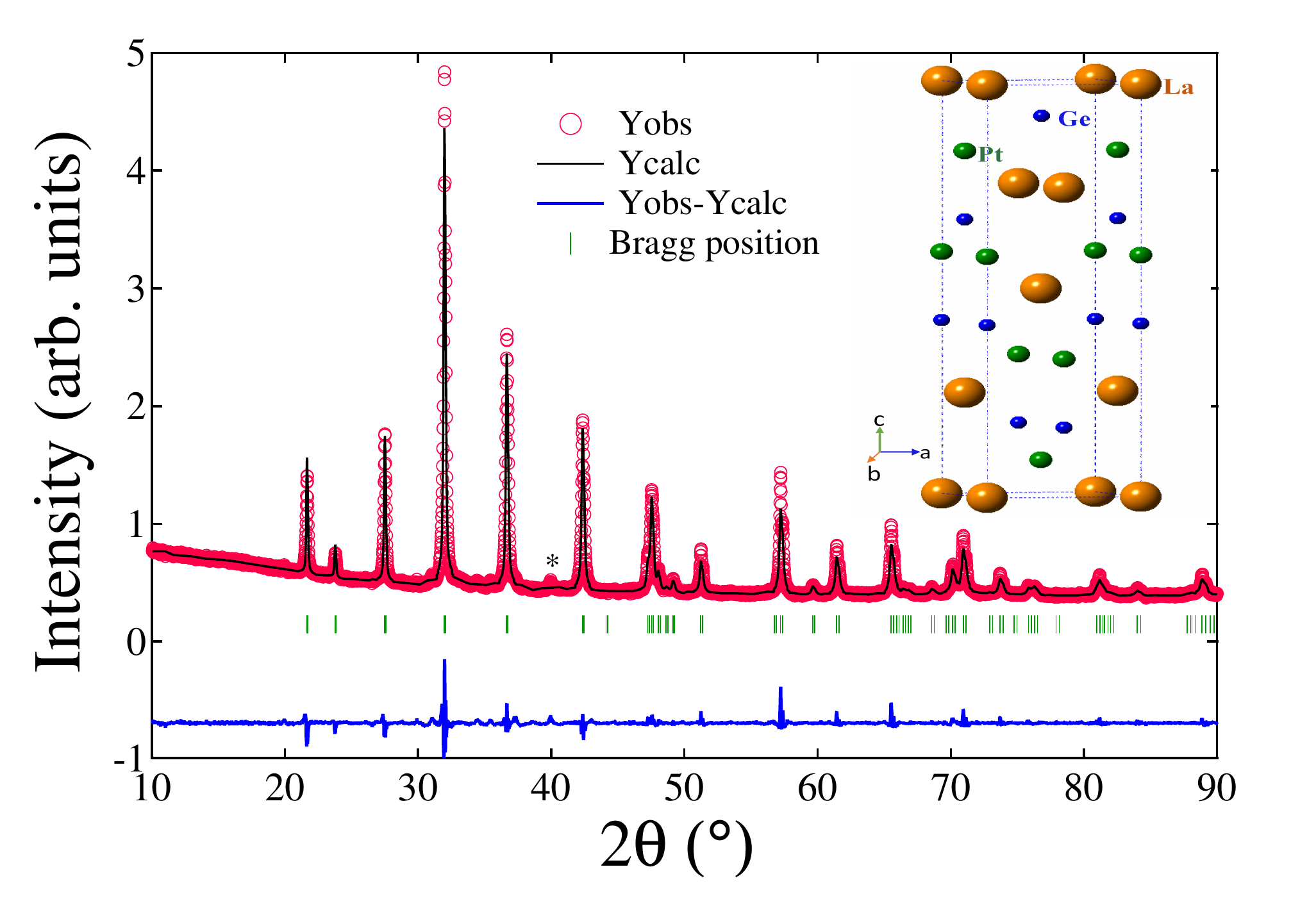}
\caption{Powder x-ray diffraction pattern for LaPtGe sample obtained at room temperature using Cu K$ _{\alpha} $ radiation (red line). The black solid line shows the Reitveld refinement whereas the blue line shows the difference between observed and calculated one. The inset shows the crystal structure.}
\label{fig1}
\end{figure}
                                                     
\section{Results and Discussion}
\subsection{Sample characterization}
The x-ray diffraction pattern is shown in Fig. \ref{fig1}. Rietveld refinement to the data was done using Fullprof software which shows the sample crystallizes into a tetragonal, noncentrosymmetric structure (space group I4$ _{I} $md ) with derived lattice parameters a = b = 4.2655(2) \text{\AA}, c = 14.9654(1) \text{\AA}. The lattice parameters obtained in this work are in good agreement with data reported previously \cite{J}. A small impurity peak is observed around 40$\textdegree$ (denoted by an asterisk) due to Pt$ _{3} $Ge. Any significant effect of this impurity phase is not observed in bulk and muon spectroscopy measurements. The inset in Fig. \ref{fig1}  shows the crystal structure of LaPtGe.

\subsection{Normal and superconducting state properties}
\subsubsection{Electrical resistivity}
\begin{figure} 
\includegraphics[width=1.0\columnwidth, origin=b]{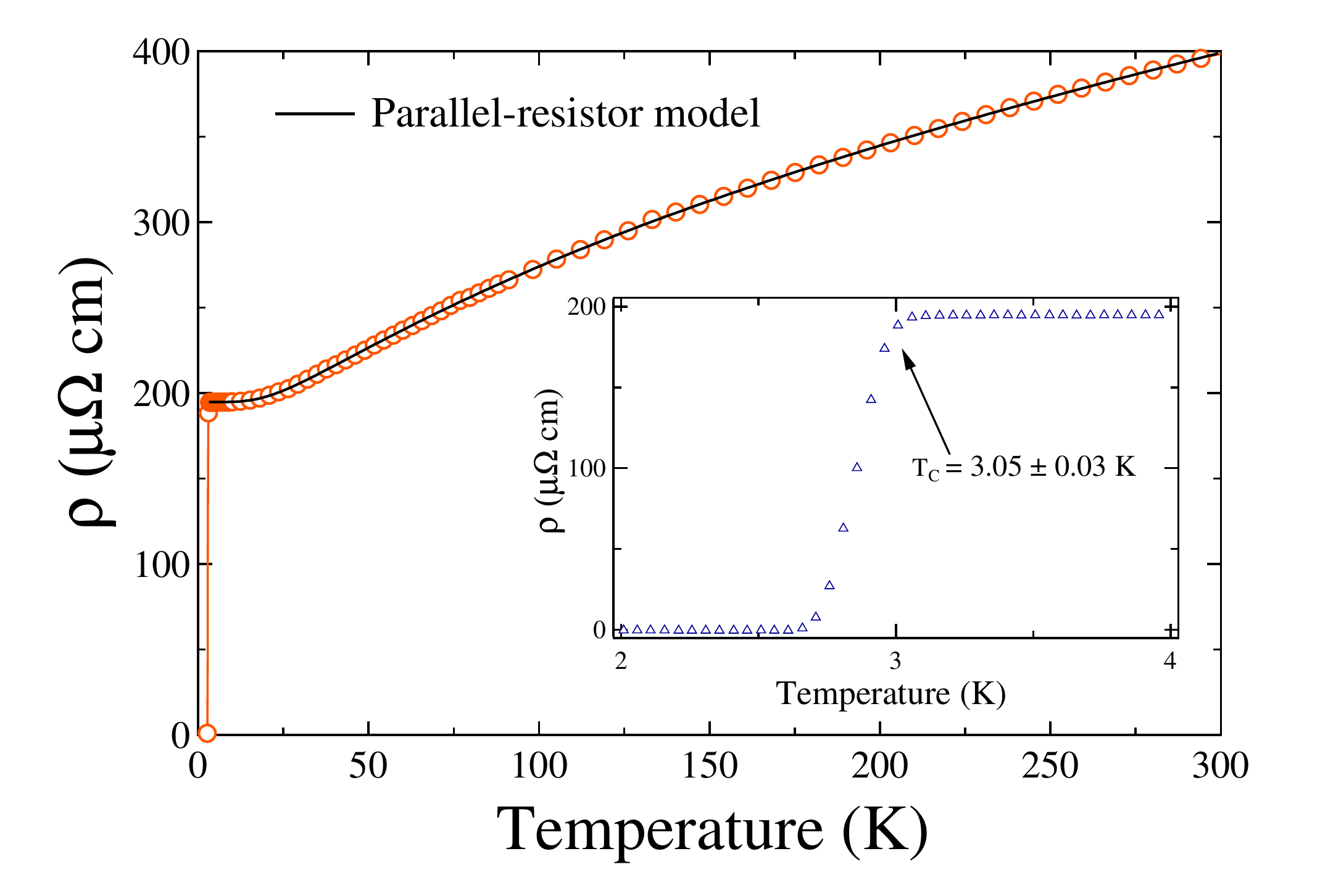}
\caption{Temperature dependence of resistivity in the range 1.8 K $ \leq $ T $ \leq $ 300 K. The inset shows the drop in resistivity at the superconducting transition, T$ _{C} $ = 3.05 $\pm$ 0.03 K. The normal state resistivity fitted with the parallel-resistor model in the temperature range 5 K $ \leq $ T $ \leq $ 300 K.}
\label{fig2}
\end{figure}

Temperature dependence of electrical resistivity $\rho(T)$ of LaPtGe in the temperature range 1.8 K to 300 K in zero applied magnetic field is shown in Fig. \ref{fig2}. A characteristic drop in resistivity, observed at $ T_{C} $ = 3.05 $\pm$ 0.03 K is shown in the inset of Fig. \ref{fig2}. The residual resistivity ratio is comparable to other LaNiSi structure compounds \cite{LPS}. A high value of absolute resistivity with the saturation behaviour at high temperatures indicate that the data can well be described by parallel-resistor model \cite{HWM}. The saturation at high temperatures typically happens when the apparent mean free path becomes short, at the order of few interatomic spacing \cite{ZFG,Para}. In such a scenario, the scattering cross section will not be linear in scattering perturbation. At high temperatures, the dominant temperature dependent scattering mechanism is electron-phonon interaction. So the resistivity will not be proportional to the mean square atomic displacement which is proportional to T for harmonic potential. Instead, it will rise less rapidly with T showing a saturating behaviour. According to this model electrical resistivity, for  T $>$ T$ _{C} $, is given by \cite{HWM,RZ2}:

\begin{equation}
 \rho(0) = \left[\frac{1}{\rho_{s}} + \frac{1}{\rho_{i}(T)} \right]^{-1}
\label{para1}
\end{equation} 

where $ \rho_{s} $ is the temperature independent saturation value of resistivity which will attain at high temperatures. The value of $ \rho_{i}(T) $ can be written as

\begin{equation}
 \rho_{i}(T) = \rho_{i,0} + \rho_{i,L}(T)
\label{para2}
\end{equation}
 \begin{figure*}[t]
\includegraphics[width=2.0\columnwidth,origin=b]{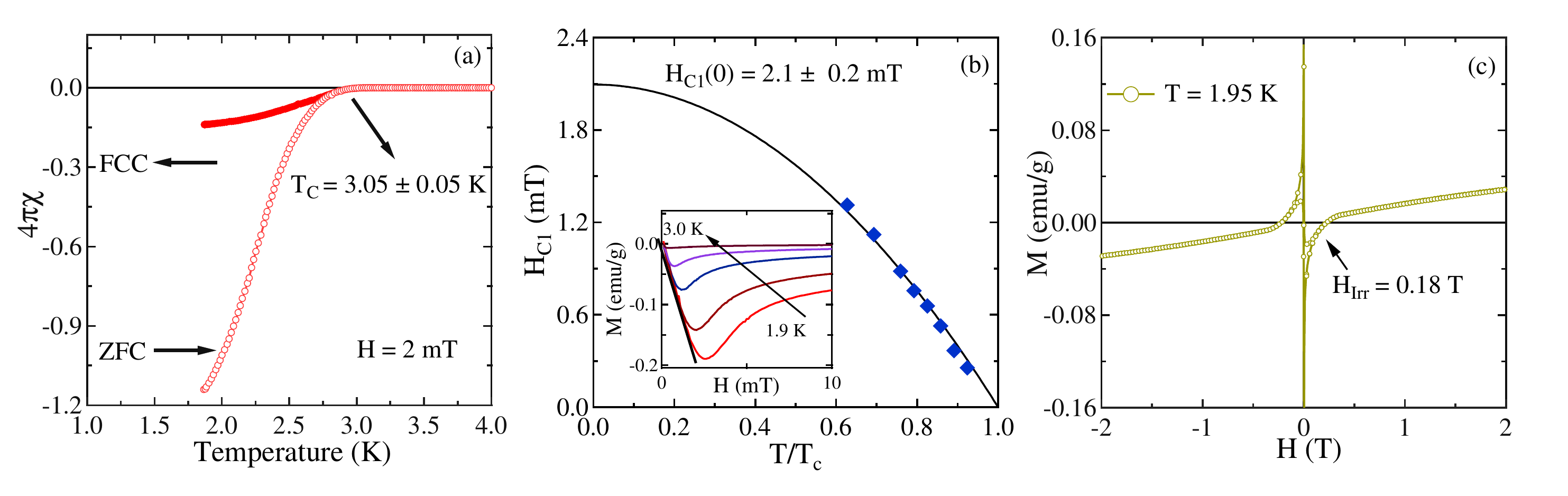}
\caption{(a) Temperature dependence of magnetic moment collected via zero field cooled warming (ZFCW) and field cooled cooling (FCC) methods under applied field of 2 mT. (b) Temperature dependence of lower critical field H$ _{C2} $. The inset shows the low field magnetization data at different temperatures. (c) Magnetization data collected at 1.95 K in the range -2 T $ \leq $ H $ \leq $ +2 T showing an irreversible field H$_{\mathrm{Irr}} $ = 0.18 mT.}
\label{fig3}
\end{figure*}
In this relation temperature independent residual resistivity which arises from impurities and disorder is accounted to $ \rho_{i,0} $, while the second term adds the temperature dependent general resistivity which can be written according to generalized Bloch-Gruneisen expression \cite{BG}

\begin{equation}
 \rho_{i,L}(T) = C\left(\frac{T}{\Theta_{D}}\right)^{5}\int_{0}^{\Theta_{D}/T}\frac{x^{5}}{(e^{x}-1)(1-e^{-1})}dx
\label{para3}
\end{equation}

where $ \Theta_{D} $ is the Debye temperature obtained from resistivity measurements, while C is a material dependent pre-factor. A fit employing this model is shown in Fig. \ref{fig2}, yields Debye temperature $ \Theta_{D} $ =  139 $ \pm $ 3 K, C = 936 $ \pm $ 7 $   \mu\Omega $cm, residual resistivity $ \rho_{0} $ = 253 $ \pm $ 2 $ \mu\Omega $cm, $ \rho_{s} $ = 848 $ \pm $ 4 $ \mu\Omega $cm.

\subsubsection{Magnetization}
\begin{figure} 
\centering
\includegraphics[width=1.0\columnwidth, origin=b]{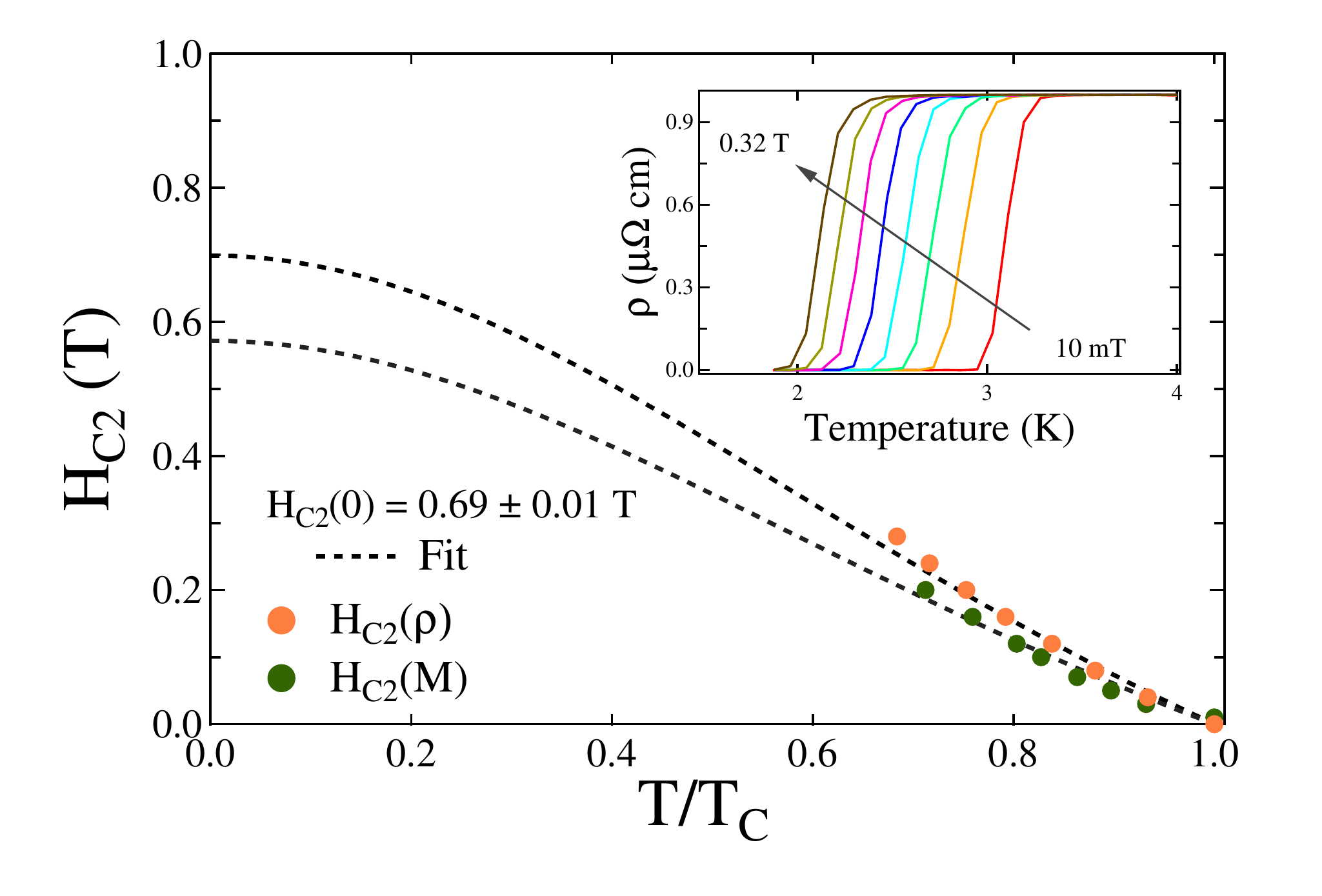}
\caption{Temperature dependence of upper critical field H$ _{C2} $ determined via magnetization and resistivity measurements. The solid lines are fit to the data using Eq. \ref{hc2}. The inset shows the resistivity curves at different applied magnetic fields. }
\label{fig4}
\end{figure}

Figure \ref{fig3}(a) shows the dc susceptibility data taken in an applied field of 2 mT in zero-field-cooled (ZFC) and field-cooled cooling (FCC) modes. Superconductivity defined as the onset of diamagnetization signal appears at T$_{c}^{onset} $ = 3.05 $\pm$ 0.05 K. The Meissner fraction value exceeds 100 \% due to uncorrected geometrical demagnetization factor. The temperature independent paramagnetic behaviour for T $>$ T$_{C}$ suggests the absence of any magnetic impurities in the sample. The low-field magnetization measurement as a function of applied magnetic field (0 to 10 mT) is taken at different temperatures to calculate the lower critical field $H_{C1}$ [see inset Fig. \ref{fig3}(b)]. The first deviation from linearity from the initial slope is taken as the basis to determine $ H_{C1}$ for all the respective temperatures. Figure \ref{fig3}(b) depicts the resulting temperature dependence of $ H_{C1} $ which is fitted by the Ginzburg-Landau equation
\\
\begin{equation}
H_{C1}(T)=H_{C1}(0)\left[1-\left(\frac{T}{T_{C}}\right)^{2}\right] .
\label{Hc1}
\end{equation} 
\\
The value of $ H_{C1} $(0) was estimated to be 2.1 $ \pm $ 0.2 mT after fitting Eq. \ref{Hc1} in $ H_{C1}(T)$ plot. Figure \ref{fig3}(c) presents the high-field magnetization loop collected at 1.95 K in the magnetic field range $\pm$2 T. The magnetic behavior exhibit conventional type-II superconductivity with an irreversible nature of magnetization below $ H_{\mathrm{Irr}} $ = 0.18 mT, above which point the applied field becomes strong enough to de-pin vortices.\\
In order to measure the upper critical field as a function of temperature $H_{C2}$(T), the shift in $T_{C}$ in different applied magnetic fields was determined from magnetization and resistivity data. Resistivity measurement as a function of temperature, $\rho$(T), was performed at different applied magnetic fields up to 0.36 T [see inset Fig. 4]. Figure \ref{fig4} displays the linear variation of  $ H_{C2} $(T) when plotted against reduced temperature t = $  T/T_{C}$. Both resistivity and magnetization is in good agreement with Ginzburg-Landau (GL) relation 
\\
\begin{equation}
H_{C2}(T) = H_{C2}(0)\left[\frac{(1-t^{2})}{(1+t^2)}\right]. 
\label{hc2}
\end{equation} 
\\
The value obtained after fitting Eq. \ref{hc2} is $ H_{C2} $(0) = 0.69 $ \pm $ 0.01 T.\\ 
Orbital limiting field $H_{C2}^{orbital}$(0) is the field where the Cooper pairs breaks due to an increased kinetic energy and is given by the Warthermar-Helfand-Hohenberg (WHH) expression \cite{EH,NRW}

\begin{equation}
H_{C2}^{orbital}(0) = -\alpha T_{C}\left.\frac{dH_{C2}(T)}{dT}\right|_{T=T_{C}}
\label{eqn4:whh}
\end{equation}
\\
where $\alpha $ is the purity factor given by 0.693 and 0.73 for superconductors in dirty and clean limit respectively. The initial slope $\frac{-dH_{C2}(T)}{dT} $ in the vicinity of $T =  T_{C}$ yields a value of 0.67 $ \pm $ 0.04  T/K , which gives $ H_{C2}^{orbital}(0)$ = 1.41 $ \pm $ 0.08 T.  Another mechanism which suppresses superconductivity is the Pauli-limiting field. According to the BCS theory, the Pauli-limiting field is given by $ H_{C2}^{P}(0)$ = 1.86 T$_{C} $, which for $ T_{C}$  = 3.05 $\pm$ 0.05 K,  produces $ H_{C2}^{p} $(0) = 5.7 $ \pm $ 0.1 T. The Maki parameter which measures the relative strengths of the orbital and Pauli-limiting fields calculated using $\alpha_{M} = \sqrt{2}H_{C2}^{orb}(0)/H_{C2}^{p}(0)$ = 0.16 $ \pm $ 0.01. Such a small value of Maki parameter implies that the effect of Pauli limiting field is negligible.\\ 
The characteristic Ginzburg-Landau coherence length $ \xi_{GL} $ can be evaluated using the $H_{C2}$(0) value from the relation \cite{tin}
\\
\begin{equation}
H_{C2}(0) = \frac{\Phi_{0}}{2\pi\xi_{GL}^{2}} ,
\label{eqn3:up}
\end{equation} 
\\
where $ \phi_{0}$  ( = 2.07 $\times$ 10$^{-15}$ Tm$^{2}$) is the magnetic flux quantum. Using $  H_{C2}(0) $ = 0.69 $ \pm $ 0.01 T, we estimated $\xi_{GL}(0)$ = 218 $\pm $ 4 \text{\AA}. The Ginzburg-Landau penetration depth $ \lambda_{GL} $(0) can be calculated from the relation 
\\
\begin{equation}
H_{C1}(0) = \frac{\Phi_{0}}{4\pi\lambda_{GL}^2(0)}\left(\mathrm{ln}\frac{\lambda_{GL}(0)}{\xi_{GL}(0)}+0.12\right).   
\label{eqn6:ld}
\end{equation} 
\\
Using the values of $H_{C1} $(0) = 2.1 $ \pm $ 0.2 mT and $\xi_{GL}(0)$ = 218 $\pm $ 4 \text{\AA}, we calculated $\lambda_{GL}(0)$ = 5047 $\pm$ 28 \text{\AA}. Ginzburg-Landau parameter for the sample is calculated with the equation
\\
\begin{equation}
\kappa_{GL} = \frac{\lambda_{GL}(0)}{\xi_{GL}(0)} 
\label{eqn7:kgl}
\end{equation}
\\
This yields a value of $\kappa_{GL}$ = 23 $ \pm $ 1. For a type-II superconductor $\kappa_{GL}$ $\ge$ $\frac{1}{\sqrt{2}}$. Therefore, it is clear that LaPtGe is a type-II superconductor. The thermodynamic critical field H$ _{C} $ can be calculated using the relation $H_{C1}(0)H_{C2}(0) = H_{C}^2\mathrm{ln}\kappa_{GL}$, giving the value as H$ _{C} $ = 21 $ \pm $ 1 mT.

\subsubsection{Specific heat}
Specific heat data were collected in the temperature range 1.9 K $\leq $ T $ \leq$ 300 K. The bulk nature of the superconducting state is evidenced by the occurrence of a well-developed discontinuity at T$ _{C} $ = 2.9 $ \pm $ 0.05 K. The normal state specific heat data can be extracted by the relation 
\\
\begin{equation}  
\frac{C}{T}=\gamma_{n}+\beta T^{2} . 
\label{eqn3:hc}    
\end{equation} 

\begin{figure} 
\centering
\includegraphics[width=1.0\columnwidth, origin=b]{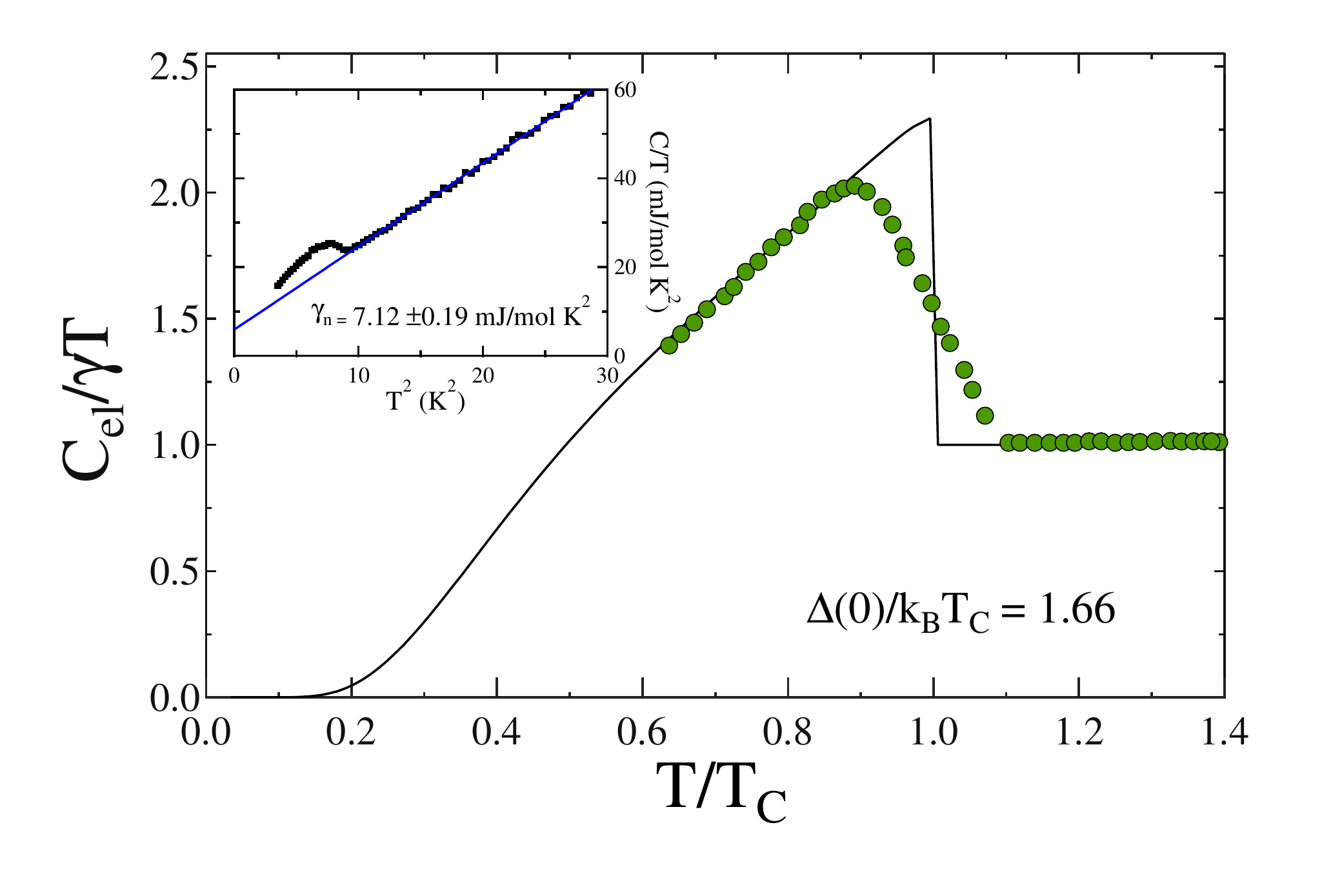}
\caption{Temperature dependence of electronic specific heat fitting with isotropic BCS expression (Eq. \ref{eqn11:Cel}). The inset shows the C/T vs T$ ^{2} $ data fit to Eq. \ref{eqn3:hc}.}
\label{fig5}
\end{figure}

Fitting the above equation in the data above T$_{C}$ determine the Sommerfeld coefficient ($ \gamma_{n} $), which describes the electronic contribution and the Debye constant ($ \beta $) representing phononic contribution. The solid blue line in the inset of Fig. \ref{fig5} in the normal state region ( 10 K $\leq T^{2} \leq$ 60 K ) represents the best fit to the data which yields  $ \gamma_{n}$ = 7.12 $\pm$ 0.19  mJ/molK$^{2}$, $\beta = 1.81 \pm 0.01$  mJ/molK$^{-4}$. The Debye temperature $\theta_{D},$ can be estimated with the relation \cite{rh}
\\
\begin{equation} 
\theta_{D}= \left(\frac{12\pi^{4}RN}{5\beta}\right)^{\frac{1}{3}}
\label{eqn4:dt}  
\end{equation}
\\
The estimated value of $\theta_{D}$ = 147 $ \pm $ 4 K is consistent with the value obtained from the parallel-resistor model. Under the assumption of a degenerate electron gas of non-interacting particles, the Sommerfeld coefficient $ \gamma_{n}$  is proportional to the density of states at the Fermi level. The value of $ \gamma_{n} $ can be used to estimate the density of states at the Fermi level $D_{C}(E _{\mathrm{F}} )$ via the relation
\\
\begin{equation} 
\gamma_{n}= \left(\frac{\pi^{2}k_{B}^{2}}{3}\right)D_{C}(E_{\mathrm{F}})
\label{eqn5:ds}  
\end{equation}
\\
where k$ _{B} $ is Boltzmann constant. Substituting $ \gamma_{n} $= 7.12 $\pm$ 0.19 mJ/mol$ K^{2} $, yields D$ _{C} $(E$ _{\mathrm{F}} $) =  3.02 $ \pm $ 0.03 $ \frac{states}{eV f.u} $. The electron-phonon coupling constant $ \lambda_{e-ph} $, a dimensionless number which describes the coupling between  electron and phonon is given by McMillans equation \cite{WL}
\\
\begin{equation}
\lambda_{e-ph} = \frac{1.04+\mu^{*}\mathrm{ln}(\theta_{D}/1.45T_{C})}{(1-0.62\mu^{*})\mathrm{ln}(\theta_{D}/1.45T_{C})-1.04 }
\label{eqn8:ld}
\end{equation}
\\
where $ \mu^{*} $  is the repulsive screened Coulomb potential having typical material specific values in the range 0.1 $ \leqslant $ $ \mu^{*} $ $ \leqslant $ 0.15, where 0.13 is used for intermetallic superconductors.  Incorporating the values of $ \theta_{D} $ and $ T_{C} $ yields $ \lambda_{e-ph} $= 0.67 $ \pm $ 0.03, which is comparable to those in noncentrosymmetric superconductors such as 0.63 in Re$_{6}$Hf \cite{rhf}  and 0.5 for  LaRhSi$_{3} $ \cite{lrs} indicating that the electron-phonon coupling is moderately strong in LaPtGe. The bare band structure density of states $ D_{band}(E_{\mathrm{F}}) $  and m*$_{band} $ are related to $ D_{C}(E_{\mathrm{F}}) $ by the relations
 \\
 \begin{equation}
 D_{C}(E_{\mathrm{F}}) = D_{band}(E_{\mathrm{F}})(1+\lambda_{e-ph})\\ 
 \label{eqn6:dk}                                                                     \end{equation}
 \begin{equation}
 m^{*} = m^{*}_{band}(1+\lambda_{e-ph})\\
 \label{eqn7:mel}                              \end{equation}
\\
Using the value of $ \lambda_{e-ph} $ = 0.67 in Eq. \ref{eqn6:dk} and  \ref{eqn7:mel} yields $ D_{band} $ (E$_{\mathrm{F}}$) = 1.81 $ \pm $ 0.06 $ \frac{states}{eV f.u} $ and the effective mass of quasiparticle as 1.67m$ _{e} $ where we used $ m^{*}_{band} $ = m$ _{e} $. The condensation energy U(0), which is the difference between the ground state energies of the normal state and the superconducting state, can be estimated using the relation U(0) = $\frac{1}{2}\Delta^{2}(0)D_{band}(E_{\mathrm{F}})$ employing $ \Delta^{2}(0)  = 6.65\times 10^{-23}$ J and $ D_{band}(E_{\mathrm{F}}) = 1.1378\times 10^{-43}$ J$^{-1}$ mol$^{-1}$  from specific heat measurements to give U(0) = 25.1 $ \pm $ 0.4 mJ/mol.	 
\\
\begin{figure} 
\centering
\includegraphics[width=1.0\columnwidth, origin=b]{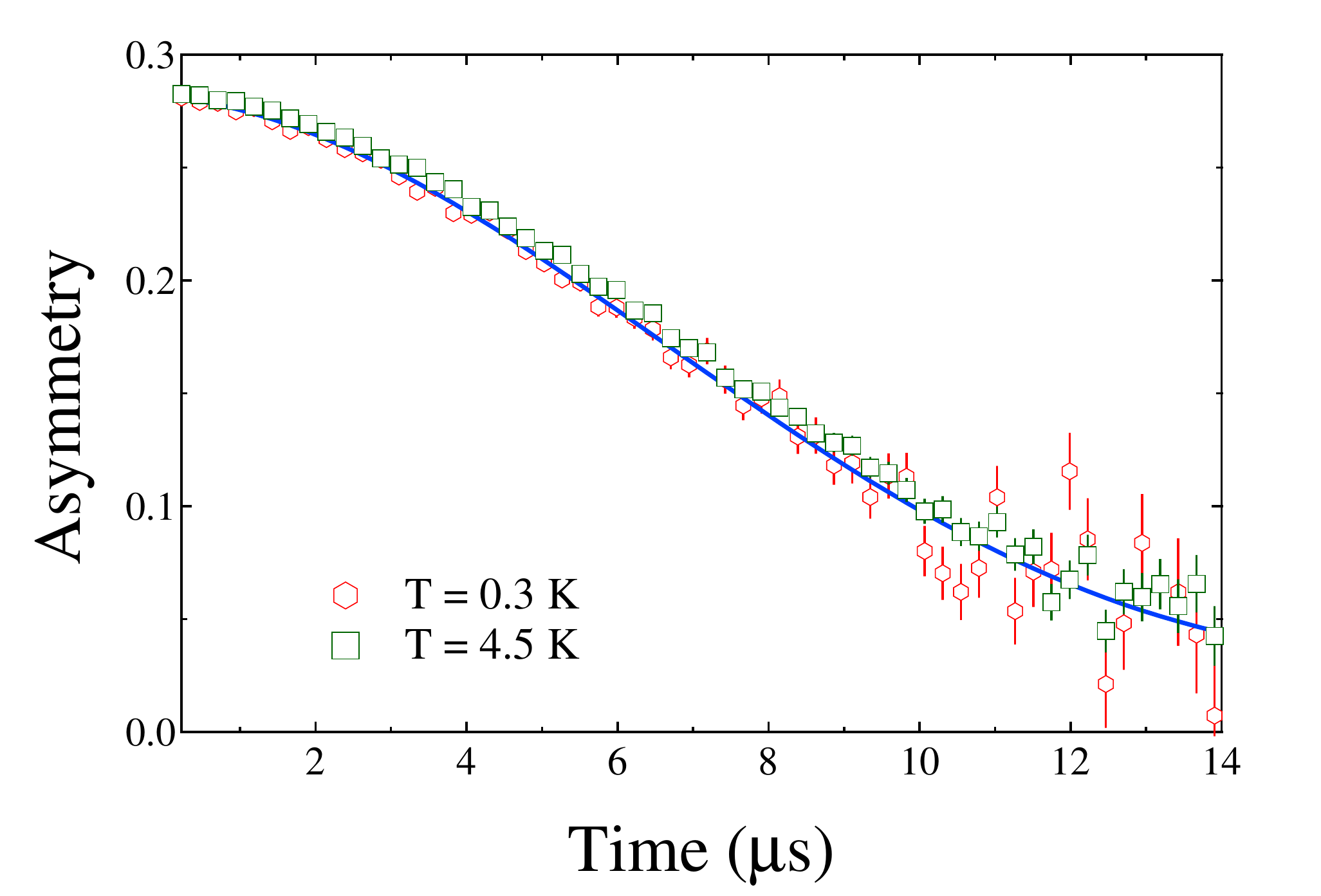}
\caption{ ZF - $ \mu $SR spectra collected above (T = 4.5 K) and below (T=0.3 K) the transition temperature. The solid line is fit to Eq. \ref{eqn18:tay}.}
\label{fig6}
\end{figure}
The electronic contribution to the specific heat, C$ _{el} $(T), is calculated by subtracting the phononic contribution of the specific heat from the total specific heat C(T).  The specific heat  jump $\frac{\Delta C_{el}}{T_{C}}$  at T$ _{C} $ is 9.2 $ \pm $ 0.5 mJ/molK$ ^{2} $. The normalized jump in specific heat is then obtained as  $\frac{\Delta C_{el}}{\gamma_{n}T_{C}}$ = 1.3 $ \pm $ 0.1, which is slightly lower than the BCS value  $\frac{\Delta C_{el}}{\gamma_{n}T_{C}}$ = 1.43. The temperature dependence of the specific heat in the superconducting state can be best described by a single gap BCS expression for normalized entropy, S
\\
\begin{equation}
\frac{S}{\gamma_{n}T_{C}} = -\frac{6}{\pi^2}\left(\frac{\Delta(0)}{k_{B}T_{C}}\right)\int_{0}^{\infty}[ \textit{f}\ln(f)+(1-f)\ln(1-f)]dy \\
\label{eqn10:s}
\end{equation}
\\
where  $\textit{f}$($\xi$) = [exp($\textit{E}$($\xi$)/$k_{B}T$)+1]$^{-1}$ is the Fermi function, $\textit{E}$($\xi$) = $\sqrt{\xi^{2}+\Delta^{2}(t)}$ , where E($ \xi $) is the energy of the normal electrons measured relative to Fermi energy, $\textit{y}$ = $\xi/\Delta(0)$, $\mathit{t = T/T_{C}}$ and $\Delta(t)$ = tanh[1.82(1.018(($\mathit{1/t}$)-1))$^{0.51}$] is the BCS approximation for the temperature dependence of energy gap. The normalized electronic specific heat is related to the normalized entropy by 
\\
\begin{equation}
\frac{C_{el}}{\gamma_{n}T_{C}} = t\frac{d(S/\gamma_{n}T_{C})}{dt} \\
\label{eqn11:Cel}
\end{equation}
\\
where $ C_{el} $ below T$ _{C} $ is described by the above equation whereas above $ T_{C} $ its equal to $ \gamma_{n}$T$_{C} $. Figure \ref{fig5} shows the fitting of the specific heat data using Eq. \ref{eqn11:Cel}.
Fitting yields a value $\alpha$ = $\Delta(0)/k_{B}T_{C}$=1.66 $ \pm $ 0.02, which is slightly less than the BCS value $ \alpha $ = 1.764. 

\begin{figure}
\includegraphics[width=1.0\columnwidth, origin=b]{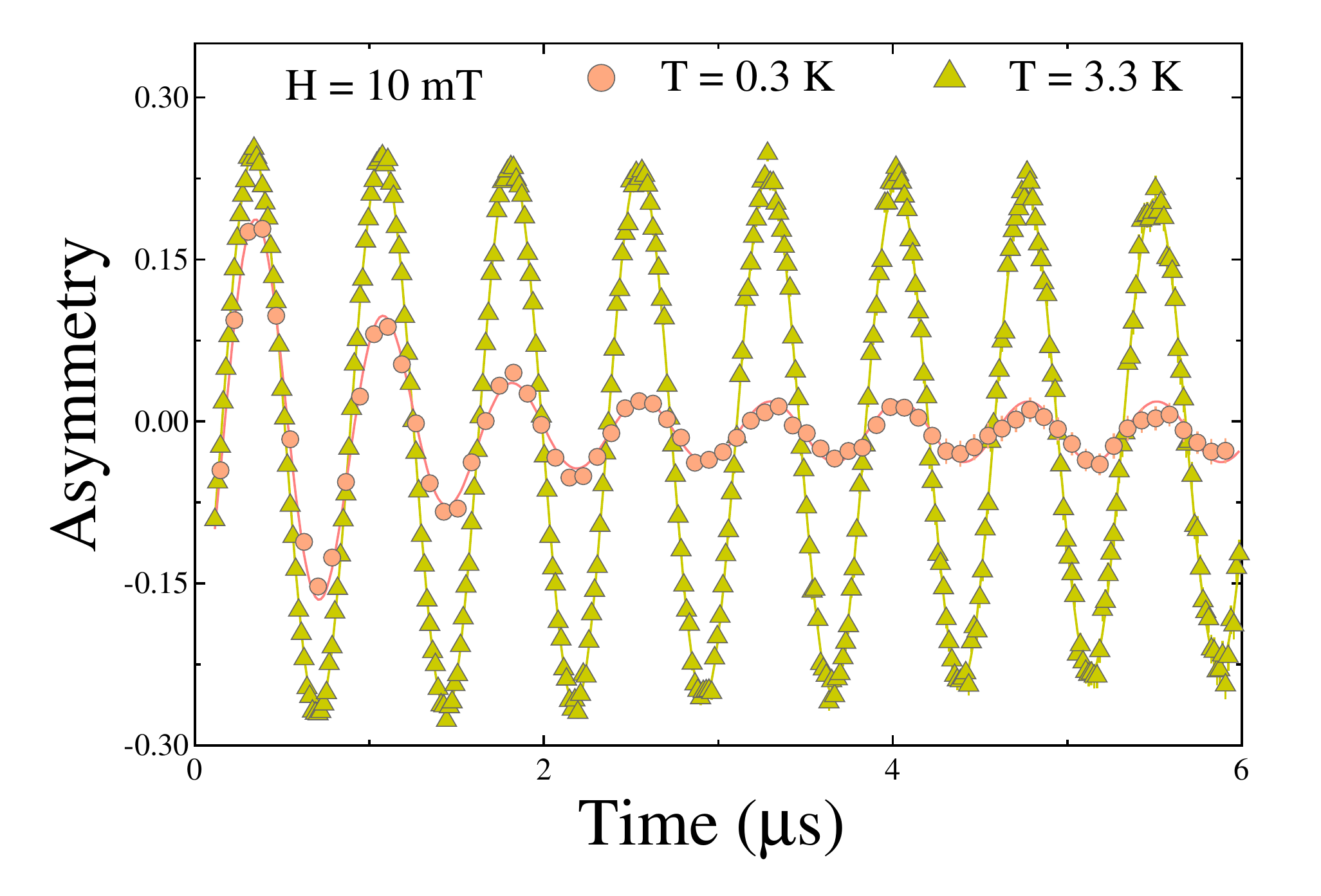}
\caption{TF - $ \mu $SR signals collected above (T = 3.3 K) and below (T = 0.3 K) the transition temperature in an applied magnetic field of 10 mT. Fast decay of signal below T$_{C}$  indicate the vortex formation. }
\label{fig7}
\end{figure}
\begin{figure*}[t]
\includegraphics[width=2.0\columnwidth,origin=b]{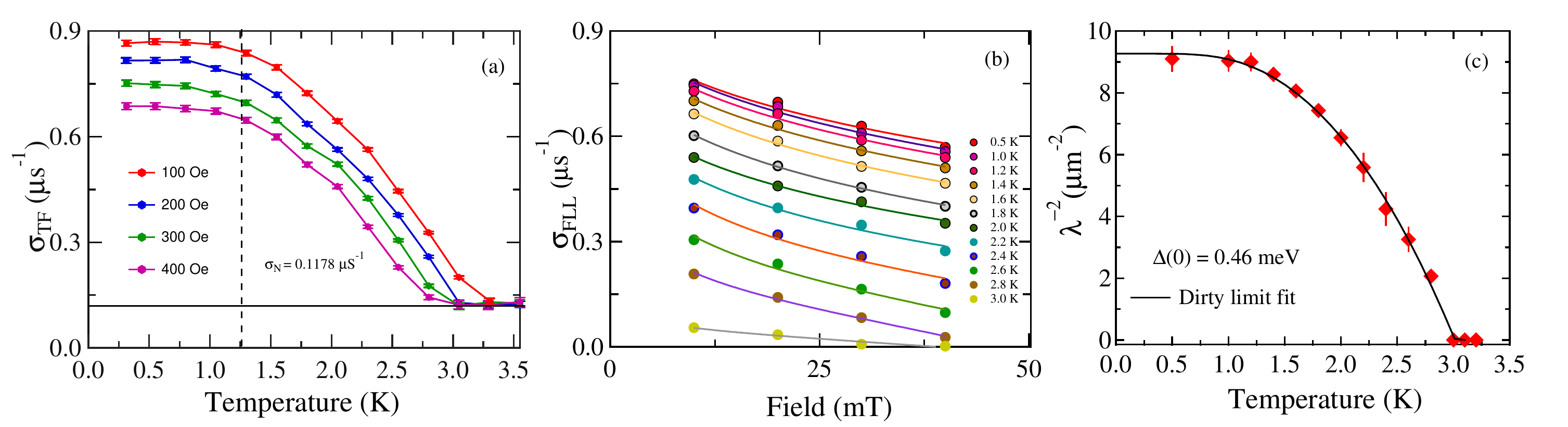}
\caption{(a) Temperature dependence of TF - $ \mu $SR depolarisation rate collected at different fields. (b) Isothermal field dependence of depolarisation. (c) Temperature dependence of inverse magnetic penetration depth square. Solid line is the fit to the s-wave model.}
\label{fig8}
\end{figure*}
\subsubsection{Muon spin relaxation and rotation}
Further analysis of the superconducting ground state of LaPtGe was carried out by muon spin rotation and relaxation ($\mu$SR) measurements. Zero-field muon spin relaxation spectra (ZF-$\mu$SR) collected at temperatures above and below T$_{C}$ as shown in Fig. \ref{fig6}. The absence of any atomic moments associated with the magnetic structure was confirmed by the non-oscillatory nature of the spectrum within the time window of $\mu$SR. The depolarization in such cases is accounted by the presence of static, randomly oriented nuclear moments. In the absence of atomic moments, muon spin relaxation in zero field is given by Gaussian Kubo-Toyabe (KT) function \cite{RSH}
\\
\begin{equation}
G_{\mathrm{KT}}(t) = \frac{1}{3}+\frac{2}{3}(1-\sigma^{2}_{\mathrm{ZF}}t^{2})\mathrm{exp}\left(\frac{-\sigma^{2}_{\mathrm{ZF}}t^{2}}{2}\right) ,
\label{eqn17:zf}
\end{equation} 
\\
where $\sigma _{\mathrm{ZF}}$ is the relaxation due to static, randomly oriented local fields associated with the nuclear moments at the muon site. The spectra can be well described by the function
\\
\begin{equation}
A(t) = A_{1}G_{\mathrm{KT}}(t)\mathrm{exp}(-\Lambda t)+A_{\mathrm{BG}} ,
\label{eqn18:tay}
\end{equation}
\\
where $ A_{1} $ corresponds to the initial asymmetry, $ \Lambda $ is the electronic relaxation rate which fluctuates on a time scale much faster than muon time scale, and $ A_{BG} $ is time-dependent background contribution from the muons stopped in the sample holder. The temperature dependence of the fit parameters $ \Lambda $ and $ \sigma $ showed no perceptible temperature dependence above and below T$ _{C}$, indicating that time-reversal symmetry is preserved within the detection limit of $ \mu $SR for LaPtGe.

Transverse field  muon spin rotation experiments (TF-$\mu$SR) was performed in an applied field of 10 mT. Figure \ref{fig7} shows the spectra collected above and below T$_{C}$. The enhanced depolarization rate below T$_{C}$ is due to the field distribution,  formed by the flux line lattice in the mixed state of the superconductor.  The TF-$\mu$SR precession signal is well described by oscillatory decaying  Gaussian function
\\
\begin{equation}
G_{\mathrm{TF}}(t) = A_{1}\mathrm{exp}\left(\frac{-\sigma^{2}_{\mathrm{TF}}t^{2}}{2}\right)\mathrm{cos}(w_{1}t+\phi)+A_{2}\mathrm{cos}(w_{2}t+\phi) ,
\label{eqn19:Tranf}
\end{equation}
\\
where $ \omega_{1} $ and $ \omega_{2} $ are the frequencies of the muon precession signal and background respectively, $ \phi $ is the initial phase offset and  $\sigma _{\mathrm{TF}}$ is the Gaussian muon spin deplorization rate. The value of  $\sigma _{\mathrm{TF}}$ depends on the distribution of vortices in the superconducting state which causes an increase in depolarization below $ T_{C} $.  $\sigma _{\mathrm{TF}}$(T) at different applied fields in the range 10 mT $ \leq $ H $ \leq $ 40 mT was extracted using Eq. \ref{eqn19:Tranf}  as shown in Fig. \ref{fig8}(a). The temperature independent depolarization due to static fields arising from the nuclear magnetic moments $ \sigma_{N} $ adds in quadrature to the contribution from the field variation across the flux line lattice $ \sigma_{\mathrm{FLL}} $.
\\
\begin{equation}
\sigma^{2}_{\mathrm{TF}} = \sigma_{\mathrm{N}}^{2}+\sigma_{\mathrm{FLL}}^{2} .\label{eqn19:sigma}
\end{equation}
\\
Field dependence of the depolarization rate $ \sigma $(H) was determined by making isothermal cuts to the $\sigma _{\mathrm{TF}}$(T) and is shown in Fig. \ref{fig8}(b). According to Ginzburg-Landau theory which explains Abrikosov hexagonal lattice in type-$\amalg$ superconductor, the magnetic penetration depth $ \lambda $ is related to $ \sigma_{\mathrm{FLL}} $ by \cite{Bran} :
\begin{equation}
\sigma_{\mathrm{FLL}} [\mu s^{-1}] = 4.854\times 10^{4}(1-h)[1+1.21(1-\sqrt{h})^{3}]\lambda^{-2} [nm^{-2}]
\label{bran}
\end{equation}
where h = $H / H_{C2} $ is the reduced field, and $ \phi_{0} $ is the magnetic flux quantum. The resulting fits to the data are shown as solid lines in Fig. \ref{fig8}(b). The estimated value of H$ _{C2}$ obtained using Eq. \ref{bran} (not shown here) is consistent with the resistivity and magnetization measurements. The temperature dependence of  $ \lambda^{-2} $ is shown in Fig. \ref{fig8}(c) where $ \lambda^{-2} $  is assumed to be zero above T$ _{C} $. The data shows a characteristic plateau at low temperature followed by a decrease as temperature increases. The temperature dependence of the superfluid density can be calculated for an isotropic s-wave superconductor  in the dirty limit using the expression
\begin{equation}
\frac{\lambda^{-2}(T)}{\lambda^{-2}(0)} = \frac{\Delta(T)}{\Delta(0)}\mathrm{tanh}\left[\frac{\Delta(T)}{2k_{B}T}\right] ,
\label{eqn22:lpd}
\end{equation}
where  $ \Delta $(T) = $ \Delta_{0} $ tanh[1.82(1.018($\mathit{(T_{C}/T})$-1))$^{0.51}$] is the BCS approximation for the temperature dependence of the energy gap. The solid lines in Fig. \ref{fig8}(c) is the result of the fit to this model for the values of $ \lambda^{-2} $ (T).
The fit yields a value of the energy gap as $ \Delta_{0} $ = 0.46 $ \pm $ 0.01 meV which gives the BCS parameter  $ \Delta_{0} $/$ k_{B}T_{C} $ = 1.79 $ \pm $ 0.07, which is very close to BCS value of 1.76 implying moderately coupled nature of the sample. The specific heat measurement also suggested the moderately coupled superelectrons where $ \Delta_{0} $/$ k_{B}T_{C} $ = 1.66 $ \pm $ 0.02. A slight difference in the energy gap value is due to the microscopic and macroscopic nature of $ \mu $SR and specific heat respectively. So the specific heat measurement along with TF-$ \mu $SR results confirm that LaPtGe is a s-wave superconductor. 

\begin{table}[h!]
\caption{Superconducting and normal state parameters of LaPtGe}
\begin{center}
\begin{tabular}[b]{lccc}\hline\hline
Parameters& unit& LaPtGe\\
\hline
\\[0.5ex]                                  
$T_{C}$& K& 3.05\\             
$H_{C1}(0)$& mT& 2.1\\                       
$H_{C2}(0)$& T& 0.69\\
$H_{C2}^{P}(0)$& T& 5.67\\
$\xi_{GL}$& \text{\AA}& 218\\
$\lambda_{GL}$& \text{\AA}& 5047\\
$k_{GL}$& &23\\
$\Delta C_{el}/\gamma_{n}T_{C}$&   &1.3\\
$\Delta(0)/k_{B}T_{C}$&  &1.66\\
$m^{*}/m_{e}$& & 7.65\\             
n& 10$^{27}$m$^{-3}$& 2.84\\
$l$&  \text{\AA}& 25.05\\ 
$\xi_{0}$&  \text{\AA}& 59\\                      
$\xi_{0}/l$& & 2.38\\
$v_{f}$& 10$^{4}$ms$^{-1}$& 6.63\\
$\lambda_{L}$& \text{\AA}& 2756\\
$T_{C}$/$T_{F}$& &0.0027\\
\\[0.5ex]
\hline\hline
\end{tabular}
\par\medskip\footnotesize
\end{center}
\end{table}

The quasiparticle number density per unit volume and mean free path related Sommerfeld coefficient via the relation \cite{ck}
\begin{equation}
\gamma_{n} = \left(\frac{\pi}{3}\right)^{2/3}\frac{k_{B}^{2}m^{*}V_{\mathrm{f.u.}}n^{1/3}}{\hbar^{2}N_{A}}
\label{eqn14:gf}
\end{equation}
where k$_{B}$ is the Boltzmann constant, N$_{A}$ is the Avogadro number, V$_{\mathrm{f.u.}}$ is the volume of a formula unit and m$^{*}$ is the effective mass of quasiparticles. The electronic mean free path $\textit{l}$ and Fermi velocity $v_{\mathrm{F}}$ is correlated with residual resistivity by the relation
 \begin{equation}
\textit{l} = \frac{3\pi^{2}{\hbar}^{3}}{e^{2}\rho_{0}m^{*2}v_{\mathrm{F}}^{2}}
\label{eqn15:le}
\end{equation}
while the Fermi velocity $v_{\mathrm{F}}$ can be written in terms of  effective mass and the carrier density by
\begin{equation}
n = \frac{1}{3\pi^{2}}\left(\frac{m^{*}v_{\mathrm{f}}}{\hbar}\right)^{3} .
\label{eqn16:n}
\end{equation}
The dirty limit expression for the penetration depth $\lambda_{GL}$(0) is given by
\begin{equation}
\lambda_{GL}(0) = \lambda_{L}\left(1+\frac{\xi_{0}}{\textit{l}}\right)^{1/2}
\label{eqn17:f}
\end{equation}
where $\xi_{0}$ is the BCS coherence length. The  London penetration depth $\lambda_{L}$, is given by
\begin{equation}
\lambda_{L} = \left(\frac{m^{*}}{\mu_{0}n e^{2}}\right)^{1/2}
\label{eqn18:laml}
\end{equation}
 The BCS coherence length $\xi_{0}$ and the Ginzburg-Landau coherence $\xi_{GL}$(0) at T = 0 K in the dirty limit is related by the expression
\begin{equation}
\frac{\xi_{GL}(0)}{\xi_{0}} = \frac{\pi}{2\sqrt{3}}\left(1+\frac{\xi_{0}}{\textit{l}}\right)^{-1/2}
\label{eqn19:xil}
\end{equation}
Eq. \ref{eqn14:gf}-\ref{eqn19:xil} form a system of four equations which can be used to estimate the parameters m$^{*}$, n, $\textit{l}$, and $\xi_{0}$ as done in Ref.\cite{DAM}. The system of equations was solved simultaneously using the values $\gamma_{n}$ = 7.1 $ \pm $ 0.19 mJ mol$^{-1}$K$^{-2}$, $\xi_{GL}$(0) = 218 $ \pm $ 4 \text{\AA}, and $\rho_{0}$ = 253 $ \pm $ 2 $\mu$ $\Omega$-cm. The estimated values are tabulated in Table II. It is clear that $\xi_{0}$ > $\textit{l}$, indicating that LaPtGe is in the dirty limit. The estimated value of mean free path $\textit{l}$ is of the same order as observed in other noncentrosymmetric superconductors, where dirty limit superconductivity was observed \cite{rw,nr2,RZ2}. 

\begin{figure}
\includegraphics[width=1.0\columnwidth]{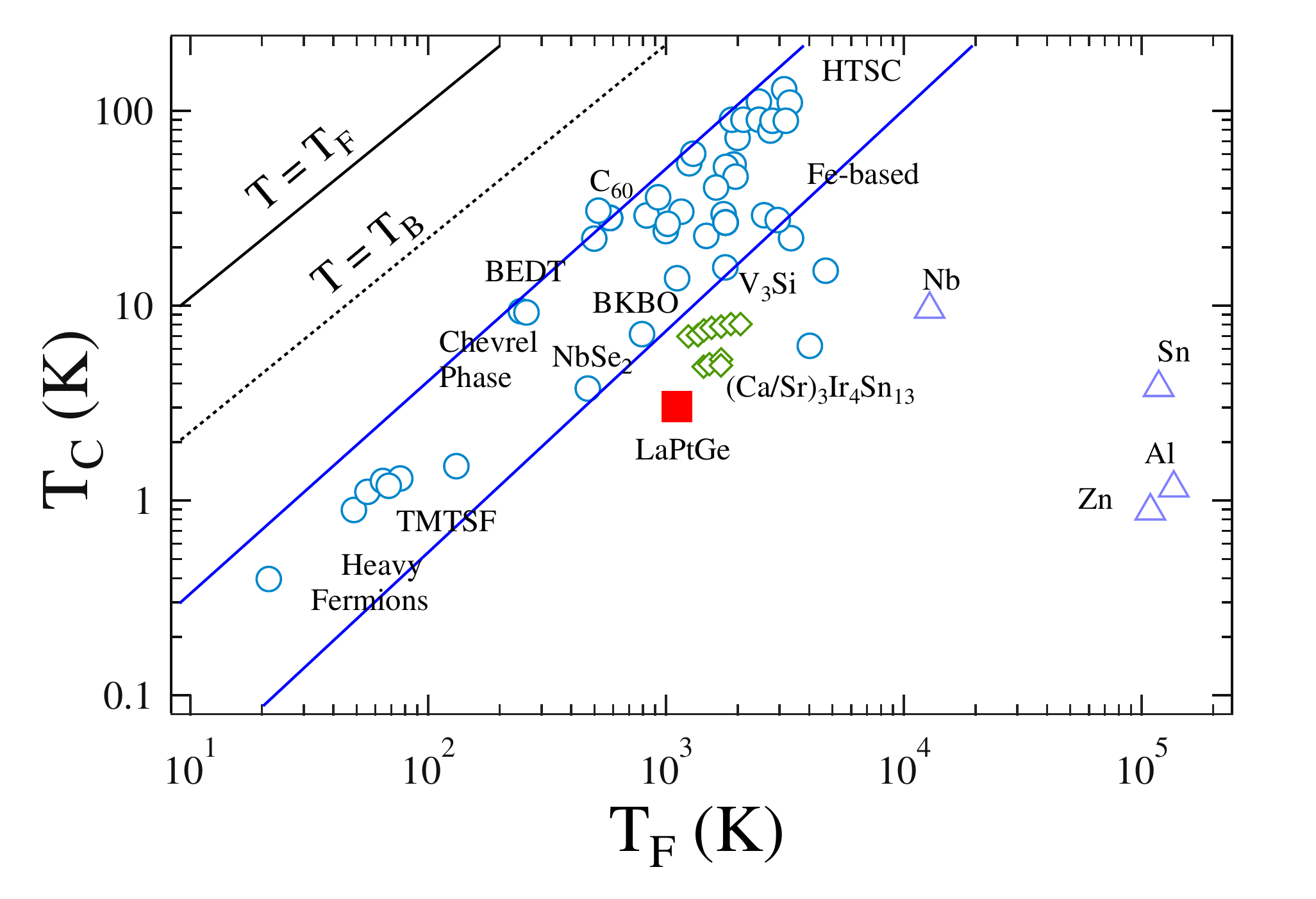}
\caption{\label{fig9} The Uemura plot showing the superconducting transition temperature $T_{c}$ vs the effective Fermi temperature $T_{F}$, where LaPtGe is shown as a solid red marker. Other data points plotted between the blue solid lines is the different families of unconventional superconductors \cite{KKC,RKH}.} 
\end{figure}
For a 3D system the Fermi temperature T$_{F}$ is given by the relation
\begin{equation}
 k_{B}T_{F} = \frac{\hbar^{2}}{2}(3\pi^{2})^{2/3}\frac{n^{2/3}}{m^{*}}, 
\label{eqn13:tf}
\end{equation}
where n is the quasiparticle number density per unit volume.

According to Uemura et al. \cite{YJU}, superconductors can be conveniently classified according to their $\frac{T_{C}}{T_{F}}$ ratio. It was shown that for unconventional superconductors this ratio falls in the range 0.01 $\leq$ $\frac{T_{C}}{T_{F}}$ $\leq$ 0.1. 
	
Using the estimated value of n in Eq. \ref{eqn13:tf} we get $T_{F}$ = 1110 K, giving $\frac{T_{C}}{T_{F}}$ = 0.0027, which places LaPtGe away from the unconventional superconductors as shown by a solid red symbol in Fig. \ref{fig9}, where solid blue lines represent the band of unconventional superconductors.\\

\section{CONCLUSION}
High purity samples of LaPtGe is prepared by arc-melting. X-ray diffraction confirm sample crystallized in noncentrosymmetric LaPtSi structure (space group no. 109). The sample exhibited superconductivity with a transition temperature T$ _{C} $ = 3.05 $ \pm $ 0.05 K. Comprehensive transport, magnetization and heat capacity measurements suggest LaPtGe is moderately coupled s-wave superconductor. Transverse field muon experiments further confirm moderately coupled s-wave superconductor. Zero-field $ \mu $SR measurements did not find any evidence of time reversal symmetry breaking in the superconducting ground state. Above mentioned results suggest the antisymmetric spin- orbital coupling is not effecting the superconducting ground state. It is clearly important to work on more noncentrosymmetric superconductors having high antisymmetric spin- orbital coupling to understand the complex superconducting ground state of these superconductors. 
\section{Acknowledgments}

R.~P.~S.\ acknowledges Science and Engineering Research Board, Government of India for the Young Scientist Grant YSS/2015/001799. We thank ISIS, STFC, UK for the muon beamtime and Newton Bhabha funding to conduct the $\mu$SR experiments.

\end{document}